\documentclass[preprint,notoc]{JHEP3}
\usepackage{epsfig}

\def\eslt{\not\!\!{E_T}}
\def\eslt{E_T^{\rm miss}}

\def\to{\rightarrow}

\def\bi{\begin{itemize}}
 \def\ei{\end{itemize}}
\def\te{\tilde e}
\def\c1p{C1^\prime}
\def\ta{\tilde a}
\def\tG{\tilde G}

\def\tu{\tilde u}

\def\ta{\tilde a}

\def\tb{\tilde b}

\def\tst{\tilde t}
\def\ttau{\tilde \tau}

\def\tg{\tilde g}

\def\tq{\tilde q}

\def\twpm{\tilde\chi^\pm}
\def\tz{\tilde\chi^0}
\def\alt{\stackrel{<}{\sim}}
\def\agt{\stackrel{>}{\sim}}
\def\be{\begin{equation}}  
\def\ee{\end{equation}}  
\def\bea{\begin{eqnarray}}  
\def\eea{\end{eqnarray}}

\title{Mainly axion cold dark matter\\
in the minimal supergravity model}
\author{Howard Baer$^{a}$, Andrew D. Box$^{a}$ and Heaya Summy$^{a}$\\
$^a$Dept.\ of Physics and Astronomy, University of Oklahoma, Norman, OK 73019, USA\\
E-mail: \email{baer@nhn.ou.edu}, \email{box@nhn.ou.edu},\email{heaya@nhn.ou.edu}}

\preprint{\vbox{}}

\abstract{
We examine the minimal supergravity (mSUGRA) model under the assumption
that the strong $CP$ problem is solved by the Peccei-Quinn mechanism.
In this case, the relic dark matter (DM) abundance 
consists of three components: 
{\it i}). cold axions, {\it ii}). warm axinos from neutralino decay, and 
{\it iii}). cold or warm thermally produced axinos. 
To sustain a high enough re-heat temperature ($T_R\agt 10^6$ GeV) 
for many baryogenesis mechanisms to function, we find that
the bulk of DM should consist of cold axions, while the admixture of
cold and warm axinos should be rather slight, with a very light axino of 
mass $\sim 100$ keV. 
For mSUGRA with mainly axion cold DM (CDM), 
the most DM-preferred parameter space regions are precisely those which 
are least preferred in the case of neutralino DM. Thus, rather
different SUSY signatures are expected at the LHC in the case of mSUGRA
with mainly axion CDM, as compared to mSUGRA with neutralino CDM. 
}

\begin{document}

\section{Introduction}
\label{sec:intro}

The cosmic abundance of cold dark matter (CDM) has been recently measured
to high precision by the WMAP collaboration\cite{wmap5}, which lately finds
\be
\Omega_{CDM}h^2=0.110\pm 0.006 ,
\ee
where $\Omega=\rho/\rho_c$ is the dark matter density relative to the 
closure density, and $h$ is the scaled Hubble constant.
No particle present in the Standard Model (SM) of particle physics 
has exactly the 
right properties to constitute CDM. However, CDM does emerge naturally 
from two compelling solutions to longstanding problems in particle physics. 

The first problem
is the strong $CP$ problem\cite{kcreview}, for which an elegant solution was 
proposed by Peccei and Quinn many years ago\cite{pq}, and which
naturally predicts the existence of a new particle\cite{ww}: 
the axion $a$. The axion turns out to be an excellent candidate 
particle for CDM in the universe\cite{absik}.

The second problem-- the gauge hierarchy problem-- arises due
to quadratic divergences in the scalar sector of the SM. 
The quadratic divergences
lead to scalar masses blowing up to the highest scale in the theory
({\it e.g.} in grand unified theories (GUTS), 
the GUT scale $M_{GUT}\simeq 2\times 10^{16}$ GeV), unless
exquisite fine-tuning of parameters is invoked. The gauge hierarchy
problem is naturally solved by introducing supersymmetry (SUSY) into the
theory. By including softly broken SUSY, 
quadratic divergences cancel between fermion and boson loops, 
and only log divergences remain. The log divergence is soft enough that
vastly different scales remain stable within a single effective theory.
In SUSY theories, the lightest neutralino emerges as an excellent
WIMP CDM candidate. The gravitino of SUSY theories is also a good super-WIMP 
CDM candidate\cite{grav}. 
Gravity-mediated SUSY breaking models include
gravitinos with weak-scale masses. These models experience tension due to
possible overproduction of gravitinos in the early universe.
In addition, late decaying gravitinos may disrupt
calculations of light element
abundances produced by Big Bang nucleosynthesis (BBN). 
This tension is known as the {\it gravitino problem}.

Of course, it is highly desirable to simultaneously account for
{\it both} the strong $CP$ problem and the gauge hierarchy problem.
In this case, it is useful to invoke supersymmetric models which
include the PQ solution to the strong $CP$ problem\cite{nillesraby}. In a 
SUSY context, the axion field is just one element of an 
{\it axion supermultiplet}. The axion supermultiplet contains 
a complex scalar field, whose real part is the $R$-parity even saxion 
field $s(x)$, and whose imaginary part is the axion field $a(x)$.
The supermultiplet also contains an $R$-parity odd spin-$1\over 2$ 
Majorana field, the axino $\ta$\cite{steffen_rev}.
The saxion, while being an $R$-parity even field, nonethless 
receives a SUSY breaking mass likely of order the weak scale. 
The axion mass is constrained by
cosmology and astrophysics to lie in a favored range 
$10^{-2}$ eV$>m_a>10^{-5}$ eV. The axino mass is very model 
dependent\cite{axmass}, and is expected to lie in the general range of keV to GeV.
An axino in this mass range would likely serve as the lightest
SUSY particle (LSP), and is also a good candidate particle for
cold dark matter\cite{rtw}.

In this paper, we investigate supersymmetric models wherein the PQ
solution to the strong $CP$ problem is also invoked. For definiteness,
we will restrict ourselves to examining the paradigm minimal
supergravity (mSUGRA or CMSSM) model\cite{msugra}. 
We will restrict our work to cases where the lightest neutralino
$\tz_1$ is the next-to-lightest SUSY particle (NLSP); the case with a
stau NLSP has recently been examined in Ref. \cite{fstw}.
Related previous work on axino DM in mSUGRA can be found in Ref. \cite{cmssm}.

We will be guided in our analysis
also by considering the possibility of including a viable mechanism
for baryogenesis in the early universe. In order to do so, we will need
to allow for re-heat temperatures after the inflationary epoch to 
reach values $T_R\agt 10^6$ GeV. We will find that in order to sustain
such high re-heat temperatures, as well as generating predominantly 
{\it cold} dark matter, we will be pushed into mSUGRA parameter
space regions that are very different from those allowed by
the case of thermally produced neutralino dark matter. In addition, we
find that very high values of the PQ breaking scale $f_a/N$ of order
$10^{11}-10^{12}$ GeV are needed, leading to the mSUGRA model with
{\it mainly axion cold dark matter}, but also with a small
admixture of thermally produced axinos, and an even smaller
component of warm axino dark matter arising from neutralino decays.
The favored axino mass value is of order 100 keV.
We note here recent work on models with dominant axion CDM explore the 
possibility that axions form a cosmic Bose-Einstein condensate, which can 
allow for the solution of several problems associated with large scale
structure and the cosmic background radiation\cite{pierre}.

The remainder of this paper is organized as follows.
In Sec. \ref{sec:disc}, we first discuss the gravitino problem, and then
examine several possible baryogenesis mechanisms: thermal and
non-thermal leptogenesis and Affleck-Dine leptogenesis.
We then examine production mechanisms for axion and thermally and 
non-thermally produced axino dark matter.
In Sec. \ref{sec:pspace}, we confront the mSUGRA model with the possibility of
mixed axion and axino cold and warm dark matter. We 
plot out contours of re-heat temperature $T_R$, and find that parameter space
regions with large enough $T_R$ to sustain at least non-thermal leptogenesis
favor a sparticle mass spectrum which is actually most
disfavored by mSUGRA with neutralino cold dark matter. 
Likewise, the regions of mSUGRA space most favored by neutralino CDM are
least favored by mixed axion/axino dark matter. This has a large impact on
the sort of SUSY signatures to be expected at LHC. The requirement
of mainly axion CDM with $T_R\agt 10^6$ GeV favors rather heavy squarks
and sleptons. Thus, we expect in this case that LHC signatures will be 
dominated by gluino pair production followed by 3-body gluino
decays to charginos and neutralinos.
In Sec. \ref{sec:conclude}, we present a summary and conclusions.

\section{The gravitino problem, leptogenesis and mixed axion/axino 
dark matter}
\label{sec:disc}

We adopt the mSUGRA model\cite{msugra} as a template model for examining the
role of mixed axion/axino dark matter in gravity-mediated SUSY
breaking models. The mSUGRA parameter space is given by
\be
m_0,\ m_{1/2},\ A_0,\ \tan\beta ,\ sign (\mu ) ,
\ee
where $m_0$ is the unified soft SUSY breaking (SSB) scalar mass 
at the GUT scale, 
$m_{1/2}$ is the unified gaugino mass at $M_{GUT}$, $A_0$ is
the unified trilinear SSB term at $M_{GUT}$ and $\tan\beta\equiv
v_u/v_d$ is the ratio of Higgs field vevs at the weak scale.
The GUT scale gauge and Yukawa couplings, and the SSB terms are
evolved using renormalization group equations (RGEs) from
$M_{GUT}$ to $m_{weak}$, at which point electroweak symmetry is
broken radiatively, owing to the large top quark Yukawa coupling. 
At $m_{weak}$, the various sparticle and Higgs boson mass matrices are 
diagonalized to find the physical sparticle and Higgs boson masses.
The magnitude, but not the sign, of the superpotential $\mu$ parameter 
is determined by the EWSB minimization conditions.
We adopt the Isasugra subprogram of Isajet for spectra generation\cite{isajet}.

\subsection{Gravitino problem}

In supergravity models, supersymmetry is broken via the superHiggs mechanism.
The common scenario is to postulate the existence of a hidden sector which
is uncoupled to the MSSM sector except via gravity. The superpotential of the 
hidden sector is chosen such that supergravity is broken, which causes 
the gravitino (which serves as the gauge particle for the superHiggs mechanism)
to develop a mass $m_{3/2}\sim m^2/M_{Pl}\sim m_{weak}$. Here, $m$ is a
hidden sector parameter assumed to be of order $10^{11}$ GeV.
\footnote{In Ref. \cite{kimnilles}, a link is suggested between 
hidden sector parameters and the PQ breaking scale $f_a$.} 
In addition to a mass
for the gravitino, SSB masses of order $m_{weak}$ are generated for
all scalar, gaugino, trilinear and bilinear SSB terms.
Here, we will assume that $m_{3/2}$ is larger than the lightest
MSSM mass eigenstate, so that the gravitino essentially decouples from
all collider phenomenology.

In all SUGRA scenarios, a potential problem arises for weak-scale
gravitinos: the gravitino problem. In this case, gravitinos $\tG$ can be
produced thermally in the early universe (even though the gravitinos
are too weakly coupled to be in thermal equilibrium) at a rate which
depends on the re-heat temperature $T_R$ of the universe. The produced $\tG$
can then decay to various sparticle-particle combinations, with a long 
lifetime of order $1-10^5$ sec (due to the Planck suppressed 
gravitino coupling constant). The late gravitino decays occur
during or after BBN, and their energy injection
into the cosmic soup threatens to destroy the successful BBN
predictions of the light element abundances.
The precise constraints of BBN on the gravitino mass and $T_R$ are
presented recently in Ref. \cite{kohri}. One way to avoid the  
gravitino problem in the case where $m_{3/2}\alt 5$ TeV is to maintain 
a value of $T_R\alt 10^5$ GeV. Such a low value of $T_R$ rules out many
attractive baryogenesis mechanisms, and so here instead we assume
that $m_{3/2}\agt 5$ TeV. In this case, the $\tG$ is so heavy that its
lifetime is of order 1 sec or less, and the $\tG$ decays near the onset of BBN.
In this case, values of $T_R$ as large as $10^9$ GeV are allowed.

In the simplest SUGRA models, one typically finds $m_0= m_{3/2}$.
For more general SUGRA models, the scalar masses are in general
non-degenerate and only of order $m_{3/2}$\cite{sugmasses}. 
Here for simplicity,
we will assume degeneracy of scalar masses, but with $m_0\ll m_{3/2}$.

\subsection{Leptogenesis}

One possible baryogenesis mechanism that requires relatively
low $T_R\sim m_{weak}$ is electroweak baryogenesis. However,
calculations of successful electroweak baryogenesis within 
the MSSM context seem to require sparticle mass spectra
with $m_h\alt 120$ GeV, and $m_{\tst_1}\alt 125$ GeV\cite{cnqw}. 
The latter requirement
is difficult (though not impossible) to achieve in the MSSM, and
is also partially excluded by collider searches for light
top squarks\cite{tev+stop}. 
We will not consider this possibility further.

An alternative attractive mechanism-- especially in light of recent
evidence for neutrino mass-- is thermal leptogenesis\cite{leptog}. 
In this scenario,
heavy right-handed neutrino states $N_i$ ($i=1-3$) 
decay asymmetrically to leptons
versus anti-leptons in the early universe. 
The lepton-antilepton asymmetry is converted to a baryon-antibaryon 
asymmetry via sphaleron effects. The measured baryon abundance can be achieved
provided the re-heat temperature $T_R$ exceeds $\sim 10^9$ GeV\cite{buchm}. 
The high $T_R$ value needed here apparently puts this mechanism into 
conflict with the gravitino problem in SUGRA theories.

A related leptogenesis mechanism called non-thermal leptogenesis
invokes an alternative to thermal production of heavy neutrinos 
in the early universe. 
In non-thermal leptogenesis\cite{NTlepto}, it is possible to have lower
reheat temperatures, since the $N_i$ may be generated via inflaton decay.
The Boltzmann equations for the $B-L$ asymmetry have been solved numerically 
in Ref. \cite{imy}.
The $B-L$ asymmetry is then converted to a baryon asymmetry via sphaleron 
effects as usual.
The baryon-to-entropy ratio is calculated in \cite{imy}, where it is found
\be
\frac{n_B}{s}\simeq 8.2\times 10^{-11}\times \left(\frac{T_R}{10^6\ {\rm GeV}}\right)
\left(\frac{2M_{N_1}}{m_\phi}\right) \left(\frac{m_{\nu_3}}{0.05\ {\rm eV}}\right) \delta_{eff} ,
\ee
where $m_\phi$ is the inflaton mass and $\delta_{eff}$ is an effective $CP$ violating phase
which may be of order 1.
Comparing calculation with data (the measured value of $n_B/s\simeq 0.9\times 10^{-10}$), 
a lower bound $T_R\agt 10^6$ GeV may be 
inferred for viable non-thermal leptogenesis via inflaton decay.

A fourth mechanism for baryogenesis is Affleck-Dine\cite{ad} 
leptogenesis\cite{my}. In this approach, a flat direction 
$\phi_i =(2H\ell_i)^{1/2}$ is identified in the scalar potential,
which may have a large field value in the early universe. When the 
expansion rate becomes comparable to the SSB terms, the field oscillates,
and since the field carries lepton number, coherent oscillations about
the potential minimum will develop a lepton number asymmetry. The lepton
number asymmetry is then converted to a baryon number asymmetry by sphalerons
as usual. Detailed calculations\cite{my} find that the
baryon-to-entropy ratio is given by
\be
\frac{n_B}{s}\simeq\frac{1}{23}\frac{|\langle H\rangle |^2  T_R}
{m_\nu M_{Pl}^2}
\ee
where $\langle H\rangle$ is the Higgs field vev, $m_\nu$ is the mass of the
lightest neutrino and $M_{Pl}$ is the Planck scale. To obtain the observed 
value of $n_B/s$, values of $T_R\sim 10^6-10^8$ are allowed for
$m_{\nu}\sim 10^{-9}-10^{-7}$ eV.

Thus, to maintain accord with either non-thermal or Affleck-Dine
leptogenesis, along with constraints from the gravitino problem, we will
aim for axion/axino DM scenarios with $T_R\sim  10^6-10^8$ GeV.

\subsection{Mixed axion/axino dark matter}

\subsubsection{Relic axions}

Axions can be produced via various mechanisms
in the early universe. Since their lifetime 
(they decay via $a\to\gamma\gamma$) turns out to
be longer than the age of the universe, 
they can be a good candidate for dark matter.
Since we will be concerned here with re-heat temperatures 
$T_R\alt 10^9\ {\rm GeV }<f_a/N$ 
(to avoid overproducing gravitinos in the early universe), 
the axion production mechanism relevant
for us here is just one: 
production via vacuum mis-alignment\cite{absik}. In this mechanism, the axion field
$a(x)$ can have any value $\sim f_a$ at temperatures $T\gg \Lambda_{QCD}$. As the temperature
of the universe drops,
the potential turns on, and the axion field oscillates and settles to its minimum 
at $-\bar{\theta} f_a/N$ (where $\bar{\theta}=\theta +arg(det\ m_q)$,
$\theta$ is the fundamental strong $CP$ violating Lagrangian parameter and $m_q$ is
the quark mass matrix).
The difference in axion field before and after potential turn-on corresponds to
the vacuum mis-alignment: it produces an axion number density
\be
n_a(t)\sim {1\over 2}m_a(t)\langle a^2(t)\rangle ,
\ee
where $t$ is  the time near the QCD phase transition.
Relating the number density to the entropy density allows one to determine the
axion relic density today\cite{absik}:
\be
\Omega_a h^2\simeq {1\over 4}\left(\frac{6\times 10^{-6}\ {\rm eV}}{m_a}\right)^{7/6} .
\label{eq:axrd}
\ee
An error estimate of the axion relic density from vacuum mis-alignment is
plus-or-minus a factor of three.
Axions produced via vacuum mis-alignment would constititute {\it cold} dark matter.
However, in the event that $\langle a^2(t)\rangle$ is
inadvertently small, then much lower values of axion relic density could be allowed.
Additional entropy production at $t>t_{QCD}$ can also lower the axion relic abundance.
Taking the value of Eq.~(\ref{eq:axrd}) literally,
and comparing to the WMAP5 measured abundance of CDM in the universe,
one gets an upper bound $f_a/N\alt 5\times 10^{11}$ GeV, or a lower bound
$m_a\agt 10^{-5}$ eV. If we take the axion
relic density a factor of three lower, then the bounds change
to $f_a/N \alt 1.2\times 10^{12}$ GeV, and $m_a\agt 4\times 10^{-6}$ eV.

\subsubsection{Axinos from neutralino decay}

If the $\ta$ is the lightest SUSY particle, then the $\tz_1$ will no longer
be stable, and can decay via $\tz_1\to \ta\gamma$.
The relic abundance of axinos from neutralino decay
(non-thermal production, or $NTP$) is given simply by
\be
\Omega_{\ta}^{NTP}h^2 =\frac{m_{\ta}}{m_{\tz_1}}\Omega_{\tz_1}h^2 ,
\label{eq:Oh2_NTP}
\ee
since in this case the axinos inherit the thermally produced
neutralino number density.
The neutralino-to-axino decay offers a mechanism to shed
large factors of relic density. For a case where $m_{\tz_1}\sim 100$
GeV and $\Omega_{\tz_1}h^2\sim 10$ (as can occur in the mSUGRA model
at large $m_0$ values)
an axino mass of less than 1 GeV reduces the DM abundance to below
WMAP-measured levels.

The lifetime for these decays has been calculated,
and it is typically in the range of $\tau (\tz_1\to \ta\gamma )\sim 0.01-1$ sec\cite{ckkr}.
The photon energy injection from $\tz_1\to\ta\gamma$ decay
into the cosmic soup occurs typically before
BBN, thus avoiding the constraints that plague the case of a gravitino LSP\cite{kohri}.
The axino DM arising from neutralino decay is generally
considered warm or even hot dark matter for cases with
$m_{\ta}\alt 1-10$ GeV\cite{jlm}.
Thus, in the mSUGRA scenario considered here, where $m_{\ta}\alt 1-10$ GeV, we usually get {\it warm} axino DM from neutralino decay.

\subsubsection{Thermal production of axinos}

Even though axinos may not be in thermal equilibrium in the early universe, 
they can still be produced thermally via scattering and decay processes in the cosmic soup.
The axino thermally produced (TP) relic abundance has been
calculated in Ref. \cite{ckkr,steffen}, and is given in Ref. \cite{steffen} using
hard thermal loop resummation as
\be
\Omega_{\ta}^{TP}h^2\simeq 5.5 g_s^6\ln\left(\frac{1.211}{g_s}\right)
\left(\frac{10^{11}\ {\rm GeV}}{f_a/N}\right)^2
\left(\frac{m_{\ta}}{0.1\ {\rm GeV}}\right)
\left(\frac{T_R}{10^4\ {\rm GeV}}\right)
\label{eq:Oh2_TP}
\ee
where $g_s$ is the strong coupling evaluated at $Q=T_R$ and $N$ is the
model dependent color anomaly of the PQ symmetry, of order 1.
For reference, we take $g_s(T_R=10^6\ {\rm GeV})=0.932$ 
(as given by Isajet 2-loop $g_s$ evolution in mSUGRA), with $g_s$ at 
other values of $T_R$ given by the 1-loop MSSM running value.
The thermally produced axinos qualify as {\it cold} dark matter as long as
$m_{\ta}\agt 0.1$ MeV\cite{ckkr,steffen}.

In Fig. \ref{fig:fatr}, we plot the re-heat temperature needed to thermally produce various
abundances of axinos versus the Peccei-Quinn scale $f_a/N$. 
We plot values of $\Omega_{\ta}^{TP}h^2=                               
0.001$ (solid), 0.01 (dashed) and 0.1 (dot-dashed),
assuming values of $m_{\ta}=10^{-4}$ (purple), $10^{-2}$ (green) and 1 GeV (maroon).
We only plot solutions with $T_R\agt 10^2$ GeV, since for lower values of $T_R\alt 10^{2-3}$ GeV, 
Eq. \ref{eq:Oh2_TP} is expected to break down.
We see from the curves that in order to achieve $T_R$ values $\agt 10^6$ GeV, 
we will need values of $f_a/N$ on the large side: $\sim 10^{11}-10^{12}$ GeV.
We also see that the purple curves-- with lowest values of $m_{\ta}\sim 100$ keV--
give the largest $T_R$ values. Of course, from the preceeding discussion, large
values of $f_a/N$ also give more {\it axion} dark matter, independent of any other
parameters. Thus, to achieve high values of $T_R$, we will likely need to
examine scenarios with mostly axion CDM, combined with smaller amounts
of cold and warm axinos.
\FIGURE[t]{
\includegraphics[width=10cm]{fatr.eps}
\caption{A plot of the expected re-heat temperature of the universe
$T_R$ versus PQ breaking scale $f_a/N$ for $\Omega_{\ta}^{TP}h^2=
0.001$ (solid), 0.01 (dashed) and 0.1 (dot-dashed), 
and with $m_{\ta}=10^{-4}$ (purple), $10^{-2}$ (green) and 1 GeV (maroon).
(The solid purple and dot-dashed green, and also the solid green and
dot-dashed maroon lines coincide.)
}\label{fig:fatr}}

\section{Preferred mSUGRA parameters with mainly axion CDM}
\label{sec:pspace}

In this section, we generate sparticle mass spectra using the
Isasugra subprogram of the event generator Isajet\cite{isajet}. 
Isasugra performs
an iterative solution of the MSSM two-loop RGEs, and includes
an RG-improved one-loop effective potential evaluation at an optimized scale,
which accounts for leading two-loop effects\cite{haber}. Complete
one-loop mass corrections for all sparticles and Higgs boson masses
are included\cite{pbmz}.\footnote{
The case of $m_{\ttau_1}<m_{\tz_1}$ was recently examined in Ref. \cite{fstw}.
In their results, they always take $\Omega_ah^2\sim 0$. We have checked using 
the Micromegas program\cite{micro} (to calculate the stau relic density,
which is not handled by IsaReD) 
that the value of $T_R$ generated in the stau NLSP
region is always less than the corresponding values generated in the 
neutralino NLSP regions for the cases considered in this section.}

Our first results are shown in Fig. \ref{fig:Oh2vfa_tr}, where we examine
the mSUGRA point with $(m_0,m_{1/2},A_0,\tan\beta ,sgn(\mu ))$ 
$=(1000, 300,0,10,+1)$ (where all mass parameters are in GeV units).
We also take $m_t=172.6$ GeV.
For this point, the neutralino relic density computed by IsaReD\cite{isared}
is $\Omega_{\tz_1}h^2=8.9$, so the point would be excluded under the assumption 
that thermal neutralinos make up the dark matter. In frame {\it a})., 
we plot the values of
$\Omega_ah^2$, $\Omega_{\ta}^{TP}h^2$ and $\Omega_{\ta}^{NTP}h^2$ versus
$f_a/N$ under the assumption that $T_R=10^6$ (dashes), $10^7$ (solid) and
$10^8$ GeV (dot-dashed). 
We assume the axion relic density is as given by the central value of 
Eq. \ref{eq:axrd}.
We require as well that the sum
$\Omega_ah^2+\Omega_{\ta}^{TP}h^2+\Omega_{\ta}^{NTP}h^2=0.11$, {\it i.e.} that the
combination of three components of axion and axino DM saturate the WMAP central 
value. For each value of $f_a/N$, the value of $m_{\ta}$ needed to saturate the
measured DM abundance is calculated, and listed in frame {\it b}). in GeV units,
along with $m_a$ in eV units. The axion abundance is of course independent of 
$T_R$. The abundance of $\Omega_{\ta}^{TP}h^2$ is fixed mainly by requiring the 
total relic abundance saturate the measured central value, and since $T_R$ is fixed, 
this means we can compute the needed value of $m_{\ta}$. For low values of 
$f_a/N\alt 10^{11}$ GeV, the DM abundance is dominated by $\Omega_{\ta}^{TP}h^2$.
(the curves for $\Omega_{\ta}^{TP}h^2$ for all three cases of $T_R$ overlap to
within the line resolution).
But comparing with frame {\it b})., we see for almost all of this range, 
$m_{\ta}<100$ keV, meaning the bulk of axino DM is actually {\it warm}, in
contradiction to what is needed to generate large scale structure in the universe.
An exception occurs in the case of $T_R=10^6$ GeV (barely enough for
non-thermal leptogenesis), where $m_{\ta}$ moves  to values higher
than $100$ keV. 
At the highest $f_a/N\agt 3\times 10^{11}$ GeV, axion CDM dominates the 
relic abundance. In this case, $m_{\ta}$ must drop precipitously so that
$\Omega_{\ta}^{TP}h^2$, which wants to rise with increasing $f_a/N$, instead
sharply drops. The region with mainly axion CDM is robust in that it gives rise
to a consistent cosmology for all choices of $T_R$: for this case, 
the axino mass can drop below 100 keV into the warm DM region, since now
axinos will only be a small component of the DM density.
\FIGURE[t]{
\includegraphics[width=10cm]{faomgh-trfix.eps}
\caption{Axion and TP and NTP axino contributions to
dark matter density for $T_R=10^6$ GeV, $10^7$ GeV and $10^8$ GeV
versus PQ breaking scale $f_a/N$.
}\label{fig:Oh2vfa_tr}}

In Fig. \ref{fig:Oh2vfa_ma}, we plot again the same quantities 
as in Fig. \ref{fig:Oh2vfa_tr}, but this time we keep $m_{\ta}$
fixed to a value of 100 keV (solid) and 1 MeV (dashed), and we 
allow $T_R$ to vary in order to
maintain the WMAP measured abundance of CDM. Frame {\it a}). shows
the relic density of all three components of axion/axino dark matter, 
while frame {\it b}). shows the value of $T_R$ needed for each value of
$f_a/N$. We see that for low values of $f_a/N$, the value of $T_R$ is
well below the $10^6$ GeV regime, and in fact doesn't even exceed 
$10^6$ GeV for the case of $m_{\ta}=1$ MeV. In the case of $m_{\ta}=100$ keV,
$T_R$ exceeds $10^6$ GeV for $f_a/N\agt 10^{11}$ GeV, and approaches a 
maximum for the case of mainly axion dark matter.
\FIGURE[t]{
\includegraphics[width=10cm]{faomgh-manofix.eps}
\caption{Axion and TP and NTP axino contributions to
dark matter density for $m_{\ta}=100$ keV
versus PQ breaking scale $f_a/N$.
}\label{fig:Oh2vfa_ma}}

Next, we explore the mSUGRA $m_0\ vs.\ m_{1/2}$ plane for
the presence of solutions with $T_R\agt 10^6$ GeV so they yield 
consistent baryogenesis mechanisms. In our first try, we set 
$f_a/N=1.2\times 10^{12}$ GeV so that the measured dark matter
density is saturated by cold axions (we assume the factor of three
downward fluctuation in $\Omega_ah^2$, which allows for an increased value of
$f_a/N$). We will assume equal
portions of TP and NTP axinos, which saturate the $1-\sigma$ error
bars on the WMAP measured $\Omega_{CDM}h^2$ value:
$\Omega_{\ta}^{TP}h^2=\Omega_{\ta}^{NTP}h^2=0.003$.
We also adopt mSUGRA parameters $A_0=0$, $\tan\beta =10$ and $\mu >0$.
Using these values, we calculate the sparticle mass spectrum, 
$\Omega_{\tz_1}h^2$ and $m_{\tz_1}$ 
at each point in mSUGRA space. We then determine the necessary value
of $m_{\ta}$ (from $\Omega_{\tz_1}^{NTP}h^2$), and then calculate
the required value of $T_R$ (from $\Omega_{\ta}^{TP}h^2$).
We plot in Fig. \ref{fig:trtb10_eq} the color-coded regions of
$T_R$ values, along with contours of $\log_{10} T_R$. 
The lower right red region is excluded due to lack of appropriate
EWSB, while the left-side red region yields a stau NLSP. The gray
region is excluded by LEP2 limits on the chargino mass.

From Fig. \ref{fig:trtb10_eq} we see that the usual regions preferred
for neutralino cold dark matter actually give the lowest values of $T_R$:
we find $T_R<10^3$ GeV in the stau co-annihilation region and
in the hyperbolic branch/focus point (HB/FP) region.
These values may even be too small to sustain electroweak baryogenesis. 
The central blue regions of
the plot accommodate the largest values of $T_R\agt 10^4$ GeV. While yielding
a much higher $T_R$ value than the stau and HB/FP regions, even these
regions do not yield a high enough $T_R$ value to sustain non-thermal leptogenesis.
\FIGURE[t]{
\includegraphics[width=10cm]{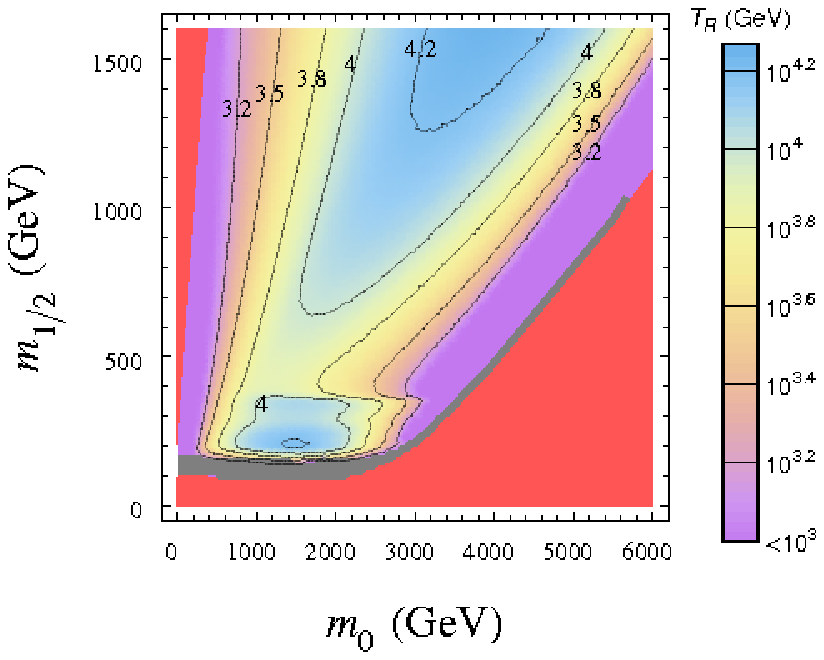}
\caption{Contours of constant $T_R$ in the $m_0\ vs.\ m_{1/2}$
plane for $A_0=0$, $\tan\beta =10$ and $\mu >0$.
We assume $\Omega_ah^2=0.11$, and 
$\Omega_{\ta}^{TP}h^2=\Omega_{\ta}^{NTP}=0.003$.
The lower right red region is excluded due to lack of appropriate
EWSB, while the left-side red region yields a stau NLSP. The gray
region is excluded by LEP2 limits on the chargino mass.
}\label{fig:trtb10_eq}}

Fig. \ref{fig:matb10_eq} shows the corresponding contours of $m_{\ta}$.
The values range from $m_{\ta}\sim 100\ (600)$ GeV in the stau (HB/FP) regions
to values of $m_{\ta}< 0.05$ GeV in the regions of high $T_R$.
\FIGURE[t]{
\includegraphics[width=10cm]{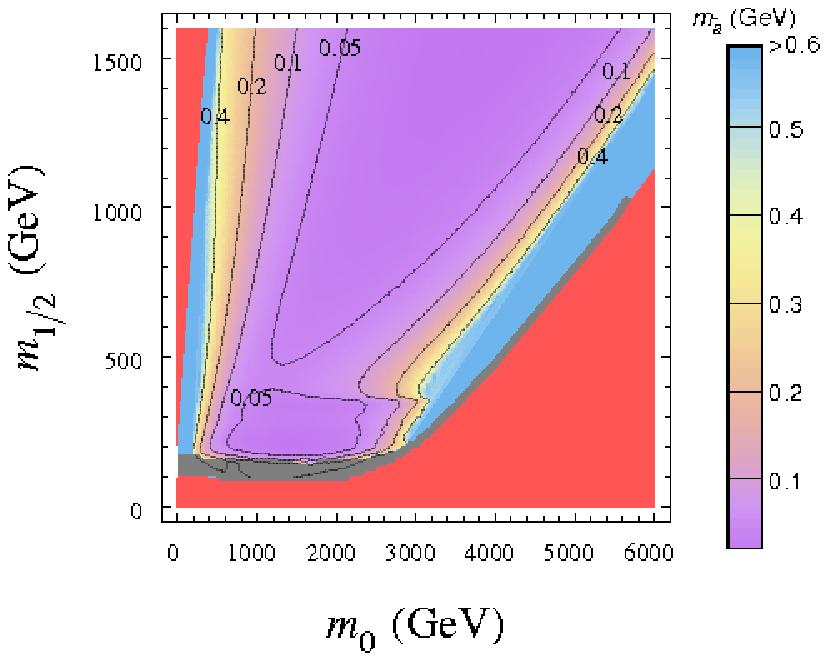}
\caption{Contours of constant $m_{\ta}$ in the $m_0\ vs.\ m_{1/2}$
plane for $A_0=0$, $\tan\beta =10$ and $\mu >0$.
We assume $\Omega_ah^2=0.11$, and 
$\Omega_{\ta}^{TP}h^2=\Omega_{\ta}^{NTP}=0.003$.
}\label{fig:matb10_eq}}
%
%
%
%
%

To push the value of $T_R$ higher, we would need to diminish even further
the value of $m_{\ta}$ (as suggested by Fig. \ref{fig:Oh2vfa_tr}{\it b}).) 
which also diminishes the amount of NTP axino dark matter. 
In Fig. \ref{fig:trtb10_neq}, we again
plot the $m_0\ vs.\ m_{1/2}$ plane for the same parameters as in
Fig. \ref{fig:trtb10_eq}
so that we saturate the CDM abundance with axions. 
However, in this case we adopt $\Omega_{\ta}^{TP}h^2=0.006$ (saturating the 
WMAP $\Omega_{CDM}h^2$ error bar), 
and take $\Omega_{\ta}^{NTP}h^2=6\times 10^{-6}$.
The much smaller value of $\Omega_{\ta}^{NTP}h^2$ 
(than that used in Fig. \ref{fig:trtb10_eq}) means the value of $m_{\ta}$
will be much smaller all over the plane than in the Fig. \ref{fig:trtb10_eq} case.
To balance the lower value of $m_{\ta}$ in the fixed value of
$\Omega_{\ta}^{TP}h^2$, a much higher value of $T_R$ will be needed.
We now see that the contours of $T_R$ plotted in Fig. \ref{fig:trtb10_neq}
move well into the $10^7$ GeV regime: enough to sustain the
non-thermal leptogenesis mechanism. In fact, the preferred regions of high
$T_R$ are precisely those regions of mSUGRA parameter space that are
{\it most disfavored} by neutralino CDM! In this case, the stau and HB/FP 
regions, preferred in mSUGRA, can only sustain $T_R$ values in the $10^3$
GeV range. The regions with $m_0\sim 1$ TeV and $m_{1/2}\sim 200-400$ GeV
allow for $T_R$ values well in excess of $10^7$ GeV. This region of parameter
space is usually neglected in simulation studies for the LHC, since it severely
disagrees with the conjecture of thermally produced neutralino CDM.
For this reason, we list a benchmark point A in Table \ref{tab:bm}
with $m_0=1500$ GeV, $m_{1/2}=200$ GeV, $A_0=0$, $\tan\beta =10$ and
$\mu >0$. In this region, squarks and sleptons have mass in the TeV range,
while $m_{\tg}\sim 500$ GeV. LHC collider events should thus be 
dominated by gluino pair production, followed by gluino three-body
decays into $q\bar{q}\tz_i$ and $q\bar{q}^\prime\twpm_i$ final states.
The $\twpm_1$ and $\tz_2$ will decay into $f\bar{f}^\prime\tz_1$ and
$f\bar{f}\tz_1$ respectively, where $f$ denotes any SM fermion states
whose decay modes are kinematically allowed. In particular, the
decays $\tz_2\to\tz_1 e^+e^-$ and $\tz_2\to\tz_1 \mu^+\mu^-$ each occur 
at $\sim 3\%$, since the decay is dominated by $Z^*$ exchange. 
\FIGURE[t]{
\includegraphics[width=10cm]{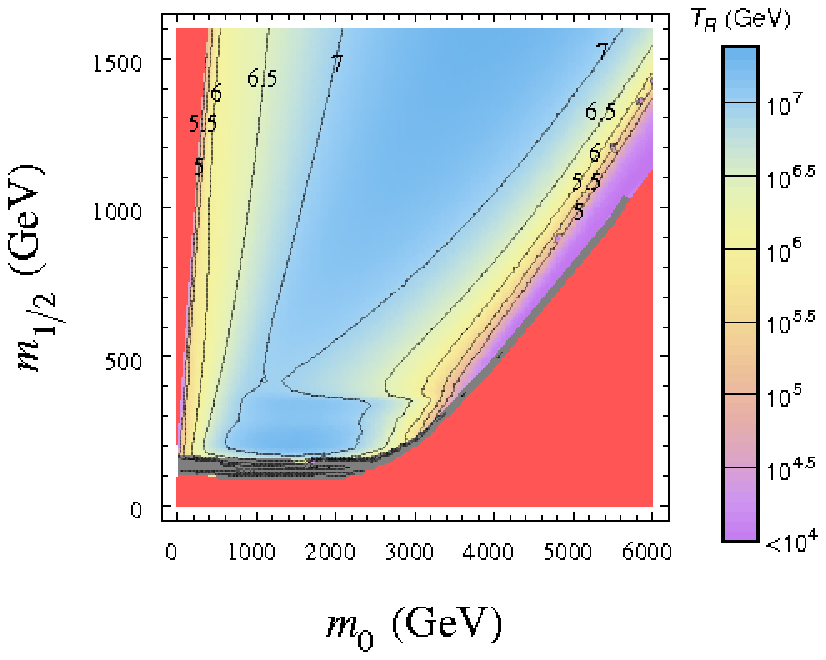}
\caption{Contours of constant $T_R$ in the $m_0\ vs.\ m_{1/2}$
plane for $A_0=0$, $\tan\beta =10$ and $\mu >0$.
We assume $\Omega_ah^2=0.11$, and 
$\Omega_{\ta}^{TP}h^2=0.006$ and $\Omega_{\ta}^{NTP}=6\times 10^{-6}$.
}\label{fig:trtb10_neq}}

Fig. \ref{fig:matb10_neq} shows contours of $m_{\ta}$ for the case as shown in 
Fig. \ref{fig:trtb10_neq}. Here, we see much smaller $m_{\ta}$ values below 
$100$ keV are generated. These low values of $m_{\ta}$
would yield warm thermally produced axinos. However, since the bulk of
DM is constituted by axions, the temperature of the small fraction of 
axinos is not relevant. 
\FIGURE[t]{
\includegraphics[width=10cm]{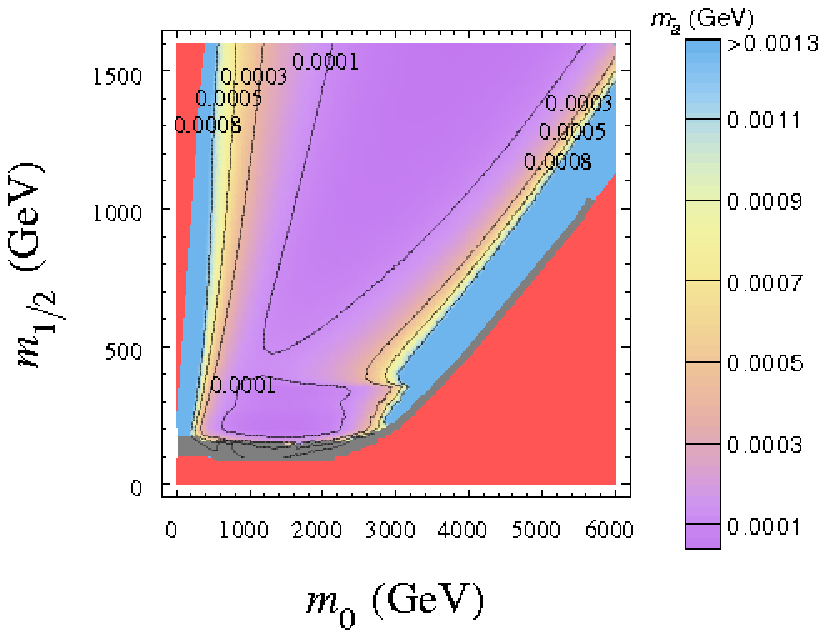}
\caption{Contours of constant $m_{\ta}$ in the $m_0\ vs.\ m_{1/2}$
plane for $A_0=0$, $\tan\beta =10$ and $\mu >0$.
We assume $\Omega_ah^2=0.11$, and 
$\Omega_{\ta}^{TP}h^2=0.006$ and $\Omega_{\ta}^{NTP}=6\times 10^{-6}$.
}\label{fig:matb10_neq}}

In Fig. \ref{fig:trtb30_neq}, we again plot contours of constant $T_R$ in the 
$m_0\ vs.\ m_{1/2}$ plane for $A_0=0$ and $\mu >0$, but this time for
$\tan\beta =30$. The larger value of $\tan\beta$ leads to larger
values of $b$ and $\tau$ Yukawa couplings, and a lower value of $m_A$\cite{ltanb}.
This in turn leads to larger rates for neutralino annihilation via
$s$-channel $A^*$ exchange diagrams, and somewhat lower neutralino 
relic density values. We again asume $\Omega_ah^2=0.11$,
$\Omega_{\ta}^{TP}h^2=0.006$ and $\Omega_{\ta}^{NTP}=6\times 10^{-6}$
and calculate the requisite value of $T_R$. From the figure, 
it is seen that the stau and HB/FP regions lead to lower values of 
$T_R\sim 10^3-10^5$ GeV, while regions with $m_0\sim 800-2000$ GeV
can still lead to $T_R$ in excess of $10^7$ GeV, 
enough to sustain non-thermal leptogenesis.  
\FIGURE[t]{
\includegraphics[width=10cm]{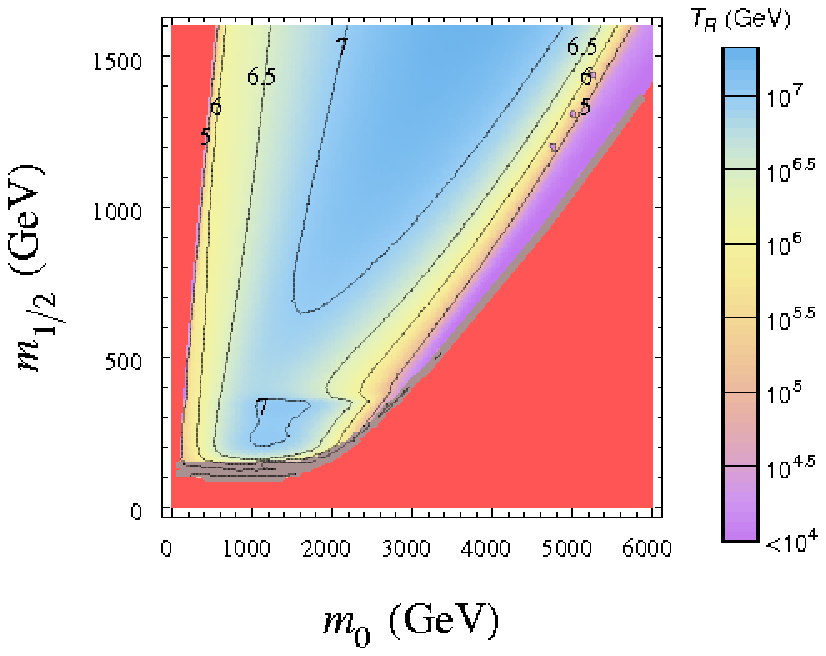}
\caption{Contours of constant $T_R$ in the $m_0\ vs.\ m_{1/2}$
plane for $A_0=0$, $\tan\beta =30$ and $\mu >0$.
We assume $\Omega_ah^2=0.11$, and 
$\Omega_{\ta}^{TP}h^2=0.006$ and $\Omega_{\ta}^{NTP}=6\times 10^{-6}$.
}\label{fig:trtb30_neq}}

The corresponding contours of $m_{\ta}$ for the $\tan\beta =30$ case are
shown in Fig. \ref{fig:matb30_neq}. They range from above 1 GeV in the 
stau co-annihilation and HB/FP region, to below 100 keV in the regions of
$T_R>10^7$ GeV. We list in Table \ref{tab:bm} an mSUGRA benchmark
point B with $\tan\beta =30$ and mainly axion CDM. 
In this case, as in the case of benchmark point A,
gluino pair production occurs at a large rate. However, in this case, 
$\tz_2\to \tz_1 Z$ at $\sim 100\%$, so LHC events will be rich
in multi-jet plus $Z$ plus $\eslt$ signatures\cite{susyz}.
\FIGURE[t]{
\includegraphics[width=10cm]{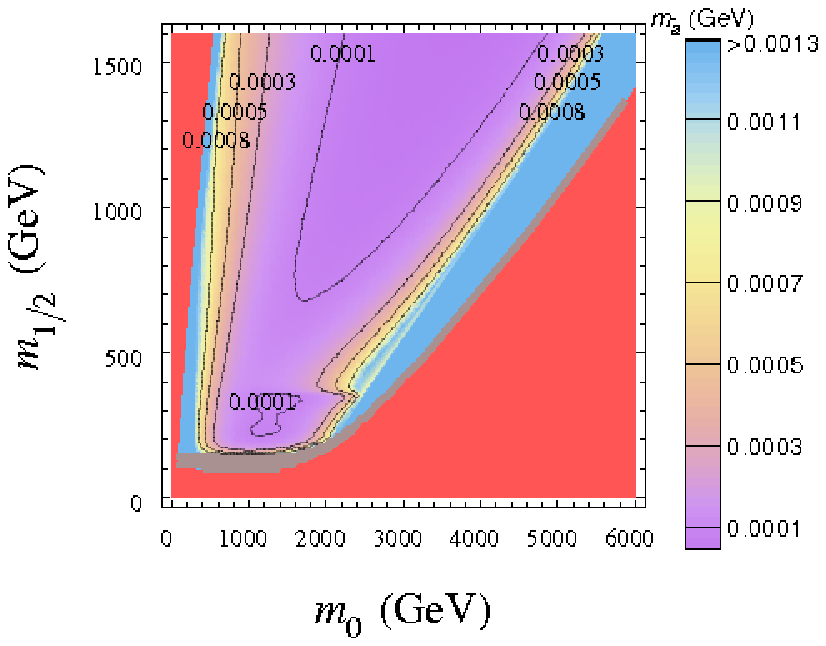}
\caption{Contours of constant $m_{\ta}$ in the $m_0\ vs.\ m_{1/2}$
plane for $A_0=0$, $\tan\beta =30$ and $\mu >0$.
We assume $\Omega_ah^2=0.11$, and 
$\Omega_{\ta}^{TP}h^2=0.006$ and $\Omega_{\ta}^{NTP}=6\times 10^{-6}$.
}\label{fig:matb30_neq}}
%

%
\begin{table}\centering
\begin{tabular}{lcc}
\hline
parameter & Pt. A & Pt. B \\
\hline
$m_0$        & 1500 & 1000 \\
$m_{1/2}$    & 200 & 300 \\
$A_0$        & 0 & 0 \\
$\tan\beta$  & 10 & 30 \\
$\mu$       & 304.5 & 368.4 \\
$m_{\tg}$   & 568.2 & 773.3 \\
$m_{\tu_L}$ & 1541.1 & 1178.4 \\
$m_{\tst_1}$& 912.3 & 774.2 \\
$m_{\tb_1}$ & 1264.5 & 965.7 \\
$m_{\te_R}$ & 1500.6 & 1005.8 \\
$m_{\twpm_1}$ & 148.7 & 227.8 \\
$m_{\tz_2}$ & 148.0 & 227.0 \\ 
$m_{\tz_1}$ & 80.0 & 122.4 \\ 
$m_A$       & 1510.6 & 912.2 \\
$m_h$       & 112.4 & 112.8 \\ 
$\Omega_{\tz_1}h^2$ & 9.2 & 6.5 \\
$BF(b\to s\gamma )$ & $3.1\times 10^{-4}$ & $2.5\times 10^{-4}$ \\
$\Delta a_\mu^{SUSY}$ & $1.5\times 10^{-10}$ & $9.7\times 10^{-10}$ \\
\hline
\end{tabular}
\caption{Masses in~GeV units and parameters
for two mSUGRA model benchmark points with mainly axion CDM: 
$\Omega_a h^2=0.11$.
We take $f_a/N=5\times 10^{11}$ GeV, along with
$\Omega_{\ta}^{TP}=0.006$ and $\Omega_{\ta}^{NTP}=6\times 10^{-6}$.
We also take $m_t=172.6$ GeV.
}
\label{tab:bm}
\end{table}

It should now be apparent that by conjecturing a large value of 
the PQ scale $f_a/N$ such that axion CDM saturates the relic density, 
then by taking decreasingly low values of axino mass, large
values of $T_R$ may be generated. To illustrate this graphically, we plot
in Fig. \ref{fig:TRsat} the value of $T_R$ required versus $m_{\ta}$
for the mSUGRA point $m_0=1000$ GeV,
$m_{1/2}=300$ GeV, $A_0=0$, $\tan\beta =10$ and $\mu >0$. 
We assume $\Omega_a h^2=0.11$, and that
$\Omega_{\ta}^{TP}h^2+\Omega_{\ta}^{NTP}=0.006$. For $m_{\ta}=0.08$ GeV,
the axino portion of the relic density is comprised almost entirely of
non-thermally produced axino DM from neutralino decay. As we decrease
$m_{\ta}$ from this value, the portion of NTP axino DM decreases, 
and since the sum is constant, the TP portion increases. Since the
value of $\Omega_{\ta}^{TP}h^2$ is proportional to $m_{\ta}$, a large
increase in $T_R$ is needed to keep pace. We see in the extreme limit,
values of $T_R$ as high as $10^{10}$ GeV can be generated (putting
us in conflict with the gravitino problem) for values of
$m_{\ta}$ as low as $10^{-8}$ GeV (far below the usual theory
expectations for $m_{\ta}$\cite{ckkr}). 
\FIGURE[t]{
\includegraphics[width=10cm]{axionsat.eps}
\caption{Plot of the value of $T_R$ needed versus
$m_{\ta}$ for the mSUGRA point with $m_0=1000$ GeV, 
$m_{1/2}=300$ GeV, $A_0=0$, $\tan\beta =10$ and $\mu >0$.
We assume $\Omega_a h^2=0.11$, and that 
$\Omega_{\ta}^{TP}h^2+\Omega_{\ta}^{NTP}=0.006$.
}\label{fig:TRsat}}
%

\section{Summary and conclusions}
\label{sec:conclude}

In this paper, we have examined the consequences for the 
mSUGRA model if dark matter is composed of an axion/axino
admixture, rather than neutralinos. We have considered this scenario
along with cosmological consequences of the gravitino problem
(which restricts $m_{3/2}>5$ TeV and re-heat temperatures $T_R\alt 10^9$ GeV) and 
leptogenesis. While thermal leptogenesis requires $T_R\agt 10^9$
GeV (in conflict with the gravitino problem), non-thermal
leptogenesis-- wherein heavy right-hand neutrino states are produced
additionally via inflaton decay-- can allow for successful baryogenesis
with $T_R\agt 10^6$ GeV. (In addition, Affleck-Dine leptogenesis
may occur at these values of $T_R$, although it may also occur at
even lower $T_R$ values.)

We explored mSUGRA parameter space for regions of high $T_R$
with three components of dark matter: dominant axion CDM, along with
small portions of thermally and non-thermally produced axino DM
(which may be either warm or cold). We find the highest values of 
$T_R$ occur in the regions of mSUGRA space which are typically
most {\it disfavored} by neutralino CDM. Likewise, the regions
of mSUGRA parameter space most favored by neutralino CDM are actually most
disfavored by mixed axion/axino DM. By combining high $T_R$ values
with fine-tuning considerations (which prefer lower values of
$m_0$ and especially $m_{1/2}$), we find mSUGRA with mainly axion
CDM prefers $m_0\sim 800-2000$ GeV, with $m_{1/2}\sim 150-400$ GeV.
There are several consequences of this scenario:
\bi
\item LHC SUSY events will be dominated by gluino pair production, 
followed by gluino three body decays to charginos and neutralinos. 
\item Current and future WIMP direct and indirect dark matter 
detection experiments will likely find  null results.
\item The ADMX\cite{admx}, or other direct axion detection experiments, stand a good
chance of finding an axion signal. The ultimate axion rate
predictions are of course model dependent.
\ei

A related scenario for sparticle spectra with mainly axion CDM
has been put forward in Refs. \cite{so10} in the context of 
$t-b-\tau$ Yukawa-unified SUSY models. These models-- expected from
simple $SO(10)$ SUSY GUT theories-- predict a spectrum of scalars in the
range of $5-15$ TeV: much higher than allowed in mSUGRA. Additionally,
Yukawa unified models require $\tan\beta \sim 50$. Thus, the 
mSUGRA model-- in contrast to Yukawa-unified models\cite{so10lhc}-- allows
for the possibility of first and second generation squark masses which
are accessible to LHC, and which would augment the gluino production rate, 
since $\tq\to q\tg$ decay is expected. In addition, mSUGRA models
with mainly axion CDM allow for lower values of $\tan\beta$ and hence
smaller values of $b$ and $\tau$ Yukawa couplings. In this case,
a lower multiplicity of $b$-quark jets is expected in LHC SUSY events.

\acknowledgments

We kindly thank Frank Steffen for his useful comments on the manuscript.
	
%

\end{document}